\documentclass[prb, aps, onecolumn, showpacs]{revtex4}
\usepackage{amssymb}
\usepackage{amsmath}
\usepackage[dvips]{graphicx}
\usepackage[english]{babel}

\begin{document}

\title{Particle-hole duality, integrability, and Russian doll BCS model}
\author{L. V. Bork$^{1,2}$, W. V. Pogosov$^{1,3,4}$}
\affiliation{$^{1}$Center for Fundamental and Applied Research, N. L. Dukhov All-Russia Research
Institute of Automatics, 127055 Moscow, Russia}
\affiliation{$^{2}$Institute for
Theoretical and Experimental Physics, 117218 Moscow, Russia}
\affiliation{$^{3}$Institute for
Theoretical and Applied Electrodynamics, Russian Academy of
Sciences, 125412 Moscow, Russia}
\affiliation{$^{4}$Moscow Institute of Physics and Technology,
Dolgoprudny, Moscow Region 141700, Russia}

\begin{abstract}
We address a generalized Richardson model (Russian doll BCS model), which is characterized by the
breaking of time-reversal symmetry.
This model is known to be exactly solvable and integrable. We point out that the Russian doll
BCS model, on the level of Hamiltonian, is also particle-hole symmetric. This implies that the same state can be expressed both in the particle and hole representations with two different sets of Bethe roots. We then derive exact relations between Bethe roots in the two representations, which can hardly be obtained staying on the level of Bethe equations. In a quasi-classical limit, similar identities for usual Richardson model, known from literature, are recovered from our results. We also show that these relations for Richardson roots take a remarkably
simple form at half-filling and for a symmetric with respect to the middle
of the interaction band distribution of one-body energy levels, since, in this special case,
the rapidities in the particle and hole representations up to the translation satisfy the same system of equations.
\end{abstract}

\author{}
\maketitle
\date{\today }

\section{Introduction}

Basic properties of conventional superconductors can be described by the microscopic
Bardeen-Cooper-Schrieffer (BCS) theory, which assumes that pairing between electrons is due to
their interaction through phonons. The simplest possible Hamiltonian, known as a reduced BCS
Hamiltonian, accounts only couplings between the spin up and spin down electrons having opposite
momenta; moreover, these couplings are supposed to be constant. Within the BCS theory, this
Hamiltonian is solved approximately by using a mean-field treatment \cite{BCS,Bogoliubov}.

It was shown by Richardson long time ago that the same Hamiltonian can be solved exactly
\cite{Rich1}. The approach to the problem, developed by Richardson, resembles a coordinate Bethe
ansatz method. The Hamiltonian eigenstates and eigenvalues are expressed through the set of
energy-like quantities (rapidities). They satisfy the set of nonlinear algebraic equations, which
are now widely referred to as Richardson equations. The system described by the reduced BCS Hamiltonian is closely related to the so-called Gaudin ferromagnet \cite{Gaudin}, for which the exact solution is also known.

Unfortunately, the resolution of Richardson equations is a formidable task, so that very few
explicit results have been obtained so far. However, in the case of a system, which contains a
limited number of pairs, the equations can be solved numerically. Nowadays, this approach is
used to investigate pairing correlations in ultrasmall metallic grains at low temperatures
\cite{Duk}.

In addition, the reduced BCS
Hamiltonian is integrable and exactly solvable through the algebraic
Bethe anzats (ABA) method \cite{Kulish,Camb,Pogosyan,Amico,Links2002}, so that
Richardson equations can be treated as Bethe
equations (BE). In Ref. \cite{Sierra}, the confromal field theory interpretation of the reduced BCS Hamiltonian was developed, which suggests interesting links between BCS theory and Laughlin states relevant for the
fractional Hall effect. The theory of the fractional Hall effect, in its turn, is also connected with the random-matrix models and growth problems \cite{Hall}.

Recently, one of us proposed a method to solve Richardson equations, which is based on the
analytical evaluations of integrals, similar to Selberg (Coulomb) integrals appearing in certain conformal field
theories \cite{Pogosov}. In this approach, when considering electron configurations with filling factors exceeding $1/2$, a special trick was used based on switching from the electron representation to the hole representation of the Hamiltonian (the existence of this symmetry in the solutions of Richardson equations was revealed and discussed before in Ref. \cite{We}). The Hamiltonian in the hole representation is also exactly solvable. If Richardson solution provides a complete set of Hamiltonian eigenstates, the same state can be expressed using either the electron (particle) representation or the hole representation. This rather trivial observation can, however, result in rather nontrivial consequences, since sets of Richardson roots and even their numbers, in general case, differ from each other in the two representations. In Ref. \cite{EPJB} the same idea was used for the analysis of the ground state energy of small-sized systems. In Ref. \cite{Bettelheim},
it was applied to facilitate computations of matrix elements between exact eigenstates of pairing Hamiltonian. In Ref. \cite{LinksP}, it was employed to re-examine $p+ip$ model and led to some new insights.

While in Refs. \cite{Pogosov,EPJB} the particle-hole symmetry was applied to the Hamiltonian eigenvalues only, in Ref. \cite{Faribault} it was used to address eigenvalues of other quantum invariants of Gaudin models. This analysis resulted in a set of nontrivial identities for the Richardson roots in the two representations.

The aim of the present article is to analyze an impact of a particle-hole symmetry for a more general model, known as a generalized Richardson model or Russian doll BCS model, proposed in Ref. \cite{Matreshka}. This model is a one-parameter generalization of Richardson model based on the inclusion of phases in pair scattering couplings, which break time-reversal symmetry. Russian doll BCS model
is also integrable by means of ABA \cite{Links,Links2}. Richardson model can be treated as a quasi-classical limit of Russian doll BCS model. Moreover, the
generalized Richardson model was shown to have other quite remarkable properties. For example, the
behavior of spectrum of model special limits (when one energy level decouples) can be described by means of cyclical renormalization group \cite{Matreshka}.

We point out that Russian doll BCS model is also characterized by the particle-hole duality. We then derive relations between sets of Bethe roots in the particle and hole representations, which can hardly be obtained staying on the level of BE. The identities of Ref. \cite{Faribault} follow from our equations, taken in the quasi-classical limit. Together with the two initial sets of BE in these representations, our relations form an overdetermined system of nonlinear equations. These results suggest that similar 'hidden' constraints for Bethe roots can be obtained for various exactly solvable and integrable models.

We also consider in a more detail a usual Richardson model. We show that the analyzed relations for Richardson roots can be cast in a remarkably simple form for some special realizations of one-body energy level distributions and at half-filling. If these levels are distributed symmetrically with respect
to the middle of the interaction band, Richardson equations in both representations do coincide,
so that sets of solutions are also the same (up to the translation of all rapidities). We then determine the correspondence between different solutions under the transformation from the particle representation to hole representation.

Apart from the general interest, our results might be
useful for the resolution of BE or computing correlations functions, since they provide additional
tools to tackle these complex problems \cite{Faribault,Ghent,Bettelheim}.

This paper is organized as follows. In Section II, we briefly discuss a derivation of generalized Richardson model and its solution by means of ABA following Refs. \cite{Links,Faddeev}.
 In Section III, we derive new relations between solutions of Bethe equations as a consequence of particle-hole symmetry of the model. We also show how Bethe equations can be obtained from similar considerations.
 In Section IV, we discuss the obtained relations in more details in the quasi-classical limit of the generalized Richardson model.
  We conclude in Section V.

\section{Generalized Richardson model}

\subsection{Preliminaries}
Cooper pairing in disordered metallic grains occurs between
time-reversed states \cite{Aleiner}. The appropriate Hamiltonian
responsible for the interaction in Cooper channel is
\begin{eqnarray}
H=\sum_{n=1}^L \varepsilon_{n} \left( b_{n+}^{\dagger }b_{n+ }+b_{n-
}^{\dagger }b_{n- }\right) -\sum_{n,n^{\prime}=1}^L g_{nn^{\prime }}
b_{n^{\prime }+ }^{\dagger }b_{n^{\prime }- }^{\dagger }b_{n- }b_{n+
}, \nonumber
\end{eqnarray}
where $n$ labels $L$ doubly-degenerate energy levels, $\pm$ refers to pairs of time-reversed
states,
while $b_{n\pm}^{\dagger }$ are creation operators for fermions at level $n$.

Eigenstates of this Hamiltonian can be classified in accordance
with the collection of blocked states. The state with the energy
$\varepsilon _{n}$ is blocked provided that it is occupied by a
single electron. This state then does not contribute to the
interaction energy. Hereafter we consider only the subspace without
singly-occupied levels.

We introduce pair creation $B_{n}^{\dagger }$ and destruction $B_{n}$ operators, defined as
$B_{n}^{\dagger } \equiv b_{n+}^{\dagger }b_{n- }^{\dagger }$ and $B_{n} \equiv b_{n-}b_{n+}$.
Their commutator reads
\begin{equation}
[B_{n}, B_{n^{\prime }}^{\dagger }]=\delta_{n, n{\prime }}(1-b_{n+}^{\dagger
}b_{n+}-b_{n-}^{\dagger }b_{n-}). \nonumber
\end{equation}
For the subspace without singly-occupied levels,
$b_{n+}^{\dagger}b_{n+}+b_{n-}^{\dagger}b_{n-}$ can be replaced by
$2B_{n}^{\dagger}B_{n} \equiv 2N_{n}$. In particular, the ground
state always belongs to this subspace.

The model with constant $g_{nn^{\prime }}$ ($g_{nn^{\prime }}=g$), known as
Richardson model, is exactly solvable\cite{Rich1} and
integrable\cite{Kulish,Camb,Pogosyan}. Actually, the same Hamiltonian, taken in the
thermodynamical limit, is also used in the BCS theory of
superconductivity, in which it is solved by the mean-field approach.

There also exists a more general integrable model\cite{Links}, which
is known as a generalized Richardson model (Russian doll BCS model).
Its Hamiltonian reads as
\begin{eqnarray}
H=2\sum_{n=1}^L \varepsilon_n N_n - g \sum_{n<n^{\prime }}^L
(e^{i\beta}B_{n^{\prime }}^{\dagger}B_n+ e^{-i\beta}B_{n} ^{\dagger}
B_{n^{\prime }}), \label{HamilGR}
\end{eqnarray}
where $\beta$ is an arbitrary angle. At $\beta = 0$, the Richardson
Hamiltonian reduces to Eq. (\ref{HamilGR}) up to the multiple of a
number operator $\sum_{n} N_{n}$. The generalized Richardson model
is well studied in the literature \cite{Matreshka,Links}. In particular, it is
known that this model can be solved by means of ABA \cite{Links}. Here we are
going to repeat crucial parts of this solution.

\subsection{Integrability}

Let us introduce pseudo-spin operators $\hat{S}_+=B^{\dagger}$,
$\hat{S}_-=B$ and $\hat{S}_z=\hat{N}-\mathbb{I}/2$. We then consider
the $R$-matrix of the form
\begin{eqnarray}
R(u)=\frac{1}{u+i\eta}(u\mathbb{I}\otimes\mathbb{I}+i\eta\mathbb{P}).
\nonumber
\end{eqnarray}
Here, as usual, $R$ acts on the tensor product of two linear spaces $V \otimes V$, $\mathbb{P}$
is permutation operator, which acts on the tensor product $x\otimes y$ of two elements from $V \otimes V$ as $\mathbb{P}(x\otimes y)=y\otimes x$. This $R$-matrix satisfies Yang-Baxter (YB)
equation
\begin{eqnarray}
R_{12}(u-v)R_{13}(u)R_{23}(v)=R_{23}(v)R_{13}(u)R_{12}(u-v).
\nonumber
\end{eqnarray}

Using $SU(2)$ spin $1/2$ representation of operators $\hat{S}_{z}$
and $\hat{S}_{\pm}$ we can write $R$ matrix in the form:
\begin{eqnarray}
R(u)=\frac{1}{u+i\eta}\left(
\begin{array}{ccc}
  u\mathbb{I}+i\eta(\hat{S}_{z}+1/2\mathbb{I}) & i\eta\hat{S}_{-},  \\
 i\eta\hat{S}_{+} & u\mathbb{I}+i\eta(-\hat{S}_{z}+1/2\mathbb{I})
  \\
 \end{array}
\right). \label{Rmatrix}
\end{eqnarray}

Using these $R$-matrices we represent monodromy matrix $T(u)$ as:
\begin{eqnarray}
T(u)=\Omega_{0}
R_{0L}(u-\varepsilon_L)...R_{02}(u-\varepsilon_2)R_{01}(u-\varepsilon_1),\label{Monodromymatrix}
\end{eqnarray}
where $\Omega$ is so called twist matrix
$\Omega=\exp(i\beta\hat{\sigma})$, $\hat{\sigma}=diag(1,-1)$.
$R_{0i}$ acts on $V_0\otimes V_i$, where $V_0$ is a so called
axillary subspace, which is $\mathbb{C}_2$ and $V_{i}$ is physical
subspace associated with $i$'th site which is also $\mathbb{C}_2$ in
the case of spin 1/2 representation. $\{\varepsilon_i\}_{i=1}^{L}$
is a given set of real parameters.

Using $T(u)$ and taking trace $Tr_0$ with respect to the axillary
subspace we can define the transfer matrix $t(u)$
\begin{eqnarray}
t(u)=Tr_0[\Omega_{0}
R_{0L}(u-\varepsilon_L)...R_{02}(u-\varepsilon_2)R_{01}(u-\varepsilon_1)],\label{transfermatrix}
\end{eqnarray}
which can be considered as an operator acting on
$V_L=\bigotimes_{i=1}^LV_i$, $V_i=\mathbb{C}_2$. $T(u)$ satisfies
another variation of YB equations:
\begin{eqnarray}
R_{12}(u-v)T_1(u)T_2(v)=T_2(v)T_1(u)R_{12}(u-v),\label{YBeq1}
\end{eqnarray}
from which one can see that
\begin{eqnarray}
[t(u),t(v)]=0,~\forall~u,v \in \mathbb{C}.\nonumber
\end{eqnarray}

The problem of finding eigenvectors and eigenvalues for $t(u)$
\begin{eqnarray}
t(u)|t,\{E_i\}\rangle=\Lambda(u,\{E_i\})|t,\{E_i\}\rangle,\label{eigen}
\end{eqnarray}
can be solved. Eigenvectors $|t,\{E_i\}\rangle$ and eigenvalues
$\Lambda(u,\{E_i\})$ of $t(u)$ are parameterized by the set of
parameters $\{E_i\}_{i=1}^{M}$, $M \leq L$. It is convenient to
rescale $t(u)$ as
\begin{eqnarray}
t(u) \mapsto
\prod_{i=1}^{L}\frac{u-\varepsilon_i+i\eta}{u-\varepsilon_i+i\eta/2}~t(u),\nonumber
\end{eqnarray}
then the explicit expression for eigenvalues of $t(u)$ can be written as
\begin{eqnarray}
\Lambda(u,\{E_i\})=e^{-i\beta}
\prod_{k=1}^{L}\frac{u-\varepsilon_k}{u-\varepsilon_k+i\eta/2}
\prod_{j=1}^{M}\frac{u-E_j/2+3i\eta/2}{u-E_j/2+i\eta/2} + e^{i\beta}
\prod_{k=1}^{L}\frac{u-\varepsilon_k+i\eta}{u-\varepsilon_k+i\eta/2}
\prod_{j=1}^{M}\frac{u-E_j/2-i\eta/2}{u-E_j/2+i\eta/2},\label{eigenexplicit}
\end{eqnarray}
where $\{E_i\}_{i=1}^{M}$ satisfies the set of equations (Bethe
equations)
\begin{eqnarray}
e^{-2i\beta}
\prod_{k=1}^{L}\frac{E_i/2-\varepsilon_k-i\eta/2}{E_i/2-\varepsilon_k+i\eta/2}
=
\prod_{j \neq i}^{M}\frac{E_i/2-E_j/2-i\eta}{E_i/2-E_j/2+i\eta}.\label{Bethe}
\end{eqnarray}

Using $t(u)$ one can construct an infinite set of commuting operators $t_{k}$, which can be obtained
expanding $t(u)$ in the powers of $u$ at infinity:
\begin{eqnarray}
t(u)= \sum_{k=1}u^{-k}t_{k},~[t_i,t_k]=0,~\forall~i,k.\nonumber
\end{eqnarray}

Eigenvalues of $t_{k}$ can be obtained using the explicit expression for
$\Lambda(u,\{E_i\})$ and expanding it also in powers of $u$ at
infinity. The form of $t_2$ is
\begin{eqnarray}
t_2\sim\left(\eta
\sin(\beta)\sum_{i=1}^{L}\varepsilon_i\hat{N}_i-\eta^2/2
\sum_{i<k}^L(e^{i\beta}\hat{S}_{+,k}\hat{S}_{-,i}+e^{-i\beta}\hat{S}_{-,i}\hat{S}_{+,k})\right)
+C\mathbb{I},\nonumber
\end{eqnarray}
where $C$ is some constant. Dropping out a trivial contribution
proportional to the constant and dividing by $\eta\sin(\beta)/2$ we
see that $t_2=H$ of the generalized Richardson model with
$g=\eta/\sin(\beta)$. Expanding $\Lambda(u,\{E_i\})$ and extracting
a term proportional to $u^{-2}$ after comparison with $t_2$ we
conclude that eigenvalues $E$ of $H$ (spectrum) are equal to
\begin{eqnarray}
E=\sum_{i=1}^M E_i+gM\cos(\beta),\label{spectrum}
\end{eqnarray}
where $\{E_i\}$ satisfies Bethe equations (\ref{Bethe}).

There is also another set of integrals of motion for $H$. They can be obtained
as values of $t(u)$ for special values of a spectral parameter.
\begin{eqnarray}
R_k=t(\varepsilon_k).\label{rk}
\end{eqnarray}
Using explicit expression (\ref{eigenexplicit}) for the eigenvalues of $\Lambda$ of $t(u)$
one can write eigenvalues $\lambda_k$ of $R_k$:
\begin{eqnarray}
\lambda_k=2e^{i\beta} \prod_{p=1,j\neq
k}^{L}\frac{\varepsilon_k-\varepsilon_p+i\eta}{\varepsilon_k-\varepsilon_p+i\eta/2}
\prod_{j=1}^{M}\frac{\varepsilon_k-E_j/2-i\eta/2}{\varepsilon_k-E_j/2+i\eta/2}.\label{eigenrk}
\end{eqnarray}

\subsection{Quasi-classical limit}

It is of interest that in the $\eta \longrightarrow 0$ limit generalized Richardson
model can be reduced to the Richardson model. On the level of BE
this limit resembles a quasi-classical one. Indeed, considering
$\sin \beta = \eta/g$ and taking $\eta
\longrightarrow 0$, BE are reduced to:
\begin{equation}
1=\sum_{n=1}^{L}\frac{g}{2\varepsilon _{n}-E_{j}}+\sum_{l=1 (\neq
j)}^{M}\frac{2g}{E_{j}-E_{l}}, \label{RichardsonElectrons}
\end{equation}
while eigenvalues of the Hamiltonian are given by Eq.
(\ref{spectrum}) at $\beta = 0$. Note that the additional contribution $gM$ in Eq. (\ref{spectrum})
is due to the exclusion of terms with coincident indices from a double sum in Eq. (\ref{HamilGR}).

\section{Particle-hole symmetry in generalized Richardson model}

It is remarkable that on the level of Hamiltonian $H$ one can use
a dual description in terms of "hole" operators. We will call the initial
representation of $H$ as "electron representation" and will use
index "$(e)$", while index "$(h)$" will be used for a dual picture.

Let us introduce creation operators for holes as
$b_{n+}^{(h)\dagger}=b_{n+ }^{(e)}$ and $b_{n- }^{(h)\dagger}=b_{n-
}^{(e)}$. Holes represent empty one-electron states. The creation
$B_{n}^{(h)\dagger }$ and destruction $B_{n}^{(h)}$ operators for
pairs of holes are $B_{n}^{(h)\dagger } \equiv b_{n+}^{(h) \dagger
}b_{n- }^{(h) \dagger }$ and $B_{n} \equiv b_{n- }^{(h)}
b_{n+}^{(h)}$. We also introduce an operator defined as
$N_{n}^{(h)}=B_{n}^{(h)\dagger }B_{n}^{(h)}$, which is a number
operator for the pairs of holes. Within the subspace without
singly-occupied levels, this operator is connected to $N_{n}^{(e)}$
through the simple relation:
\begin{equation}
N_{n}^{(e)}+N_{n}^{(h)}=1. \nonumber
\end{equation}
Next, we define a set of energies as
\begin{equation}
\varepsilon_{n}^{(h)}=-\varepsilon_{n}^{(e)}. \nonumber
\end{equation}

Using these quantities and replacing $\beta \mapsto -\beta$ we can
rewrite $H^{(e)}$ as $H^{(h)}$ up to the constant term. Both
Hamiltonians describe the same system.

It is assumed that there is one-to-one correspondence between sets of
eigenvectors and eigenvalues in both representations. Hamiltonian
$H^{(e)}$ can be obtained from the transfer matrix $t^{(e)}(u)$, which
was described above, while $H^{(h)}$ can be
obtained from the transfer matrix $t^{(h)}(u)$, where one has to use
operators and "inhomogeneities" $\varepsilon$ in $(h)$ representation.

When the number of particles is $M$, the number of holes is $L-M$.
Thus, there must be two equivalent representations for the same
eigenvector of $H$ (provided that the ABA yields a complete set of
solutions):
\begin{equation}
|t^{(e)},\{E^{(e)} \}_M\rangle\sim|t^{(h)},\{E^{(h)} \}_{L-M}\rangle.
\label{equivrepr}
\end{equation}
Here the set of $\{E^{(e)} \}_M$ satisfies the system of BE for $H^{(e)}$,
while $\{E^{(h)} \}_{L-M}$ satisfies the system of BE for $H^{(h)}$.

One can readily see that the following identity holds
\begin{eqnarray}
Tr_0[\Omega_{0}(\beta) R_{0L}^{(e)}(u,\eta)...R_{02}^{(e)}(u,\eta)R_{01}^{(e)}(u,\eta)]=
Tr_0[\Omega_{0}(-\beta) R_{0L}^{(h)}(-u,-\eta)...R_{02}^{(h)}(-u,-\eta)R_{01}^{(h)}(-u,-\eta)]
,\label{traces}
\end{eqnarray}
where $R_{0i}^{(e)}$ ($R_{0i}^{(h)}$) are $R$ matrices (\ref{Rmatrix}) written in
terms of the electron (hole) operators. This identity also implies
that
\begin{eqnarray}
t^{(e)}(\varepsilon_k^{(e)})=t^{(h)}(\varepsilon_k^{(h)}).\label{eht}
\end{eqnarray}
It is tempting to consider these identities as a manifestation of $\mathcal{C}\mathcal{P}$ symmetry of some kind.

Using different representations for eigenvectors of $H$ we see that this
gives the following relations between eigenvalues of $t^{(e)}$ and $t^{(h)}$:
\begin{eqnarray}
\Lambda(\varepsilon_k^{(e)},\eta,\beta,u,\{E^{(e)} \}_{M})=\Lambda(\varepsilon_k^{(h)},-\eta,-\beta,-u,\{E^{(h)} \}_{L-M}).\label{eigeneht}
\end{eqnarray}

Using Eq. (\ref{eigenexplicit}), we represent these relations in the explicit form as
\begin{eqnarray}
e^{-i\beta}
\prod_{k=1}^{L}(u-\varepsilon_k^{(e)})
\prod_{j=1}^{M}\frac{u-E_j^{(e)}/2+3i\eta/2}{u-E_j^{(e)}/2+i\eta/2} + e^{i\beta}
\prod_{k=1}^{L}(u-\varepsilon_k^{(e)}+i\eta)
\prod_{j=1}^{M}\frac{u-E_j^{(e)}/2-i\eta/2}{u-E_j^{(e)}/2+i\eta/2} \nonumber  \\
=e^{i\beta}
\prod_{k=1}^{L}(u+\varepsilon_k^{(h)})
\prod_{j=1}^{L-M}\frac{u+E_j^{(h)}/2+3i\eta/2}{u+E_j^{(h)}/2+i\eta/2} + e^{-i\beta}
\prod_{k=1}^{L}(u+\varepsilon_k^{(h)}+i\eta)
\prod_{j=1}^{L-M}\frac{u+E_j^{(h)}/2-i\eta/2}{u+E_j^{(h)}/2+i\eta/2}.
\label{eigenehtexplicit}
\end{eqnarray}
Let us stress that Eq. (\ref{eigenehtexplicit}) must be valid for arbitrary complex number $u$.
This equation therefore can be used to generate various identities for Bethe roots. For instance, taking residues of both sides of this equation at the points $E_j^{(e)}/2-i\eta/2$, one recovers $M$ Bethe equations for $\{E_i^{(e)}\}$
\begin{eqnarray}
e^{-2i\beta}
\prod_{k=1}^{L}\frac{E_i^{(e)}/2-\varepsilon_k^{(e)}-i\eta/2}{E_i^{(e)}/2-\varepsilon_k^{(e)}+i\eta/2}
=
\prod_{j \neq i}^{M}\frac{E_i^{(e)}/2-E_j^{(e)}/2-i\eta}{E_i^{(e)}/2-E_j^{(e)}/2+i\eta},\label{Betheel}
\end{eqnarray}
while taking residues at the points $-E_j^{(h)}/2-i\eta/2$, one obtains $L-M$ Bethe equations for $\{E_i^{(h)}\}$
\begin{eqnarray}
e^{2i\beta}
\prod_{k=1}^{L}\frac{E_i^{(h)}/2-\varepsilon_k^{(h)}+i\eta/2}{E_i^{(h)}/2-\varepsilon_k^{(h)}-i\eta/2}
= \prod_{j \neq
i}^{L-M}\frac{E_i^{(h)}/2-E_j^{(h)}/2+i\eta}{E_i^{(h)}/2-E_j^{(h)}/2-i\eta}.\label{Bethehole}
\end{eqnarray}
These two sets of BE in the electron and in hole representations are consistent with Eq. (\ref{Bethe}).

Choosing different values of $u$, one can establish various cross-relations between Bethe roots in the two representations. For instance, taking $u=\varepsilon_k^{(e)}$, we can write
\begin{eqnarray}
e^{i\beta}
\prod_{j=1}^{M}\frac{\varepsilon_k^{(e)}-E_j^{(e)}/2-i\eta/2}{\varepsilon_k^{(e)}-E_j^{(e)}/2+i\eta/2}=
e^{-i\beta}
\prod_{j=1}^{L-M}\frac{\varepsilon_k^{(h)}-E_j^{(h)}/2+i\eta/2}{\varepsilon_k^{(h)}-E_j^{(h)}/2-i\eta/2}.
\label{newBethe1}
\end{eqnarray}
This is a set of $L$ equations for $L$ variables.
Each equation is now somehow related to one-body energy level, in contrast to Bethe equations each being related to a given rapidity.

It is also possible to expand both sides of Eq. (\ref{eigenehtexplicit}) in powers of $1/u$ at $u\rightarrow \infty $. Equating prefactors in front of $1/u^{n+1}$, one can get various relations between $n$-th moments of sets $\{E_i^{(e)}\}$ and $\{E_i^{(h)}\}$. Such a relation for the first moments is equivalent to the condition that an energy of the same state both in the electron and hole representations to be the same.

\section{Particle-hole symmetry in Richardson model}

In this Section, we address some remarkable aspects of particle-hole symmetry in usual Richardson model.
The analogs of relations (\ref{newBethe1}) between Bethe roots for Gaudin models have been already reported in Ref. \cite{Faribault} (see Eq. (13) of this article). It is easy to see that by taking quasi-classical limit ($\sin \beta = \eta/g$ and $\eta
\longrightarrow 0$), these relations can be recovered directly from (\ref{newBethe1}). After some simple algebra, we express them in a nicely symmetric and simple form as ($m=1,2,\ldots,L$)
\begin{equation}
\sum_{j=1}^{M}\frac{g}{2\varepsilon
_{m}^{(e)}-E_{j}^{(e)}}+\sum_{j=1}^{L-M}\frac{g}{2\varepsilon
_{m}^{(h)}-E_{j}^{(h)}}=1. \label{EHLambdaFinal}
\end{equation}
Here $E_{j}^{(e)}$ ($E_{j}^{(h)}$) satisfy the system of $M$ ($L-M$) equations of the form (\ref{RichardsonElectrons}), in which $\varepsilon_{m}^{(e)}$ ($\varepsilon_{m}^{(h)}$) is used instead of $\varepsilon_{m}$.

\subsection{Electrostatic picture in the hole representation}

It is known that there exists an interesting electrostatic mapping for Richardson model
and some other related models \cite{Gaudin,Rich3,AmicoEl,RomanNucl,Pogosov}.
Namely, Richardson equations can be formally written as
stationary conditions for an energy of free classical particles with
electrical charges $2\sqrt{g}$ located on the plane with coordinates
given by (Re $E_{j}^{(e)}$, Im $E_{j}^{(e)}$). These particles are
subjected into an external uniform force directed along the axis of
abscissa with the strength $-2$. They are attracted to fixed
particles each having a charge $-\sqrt{g}$ and located at
$2\varepsilon _{m}^{(e)}$. Free charges repeal each other. The
interaction between the particles is logarithmic.

Richardson equations for holes correspond to
the electrostatic system, for which the distribution of one-energy
levels is mirror-imaged with respect to the zero energy. In addition, the
number of free charges is also different, while a direction of the
external force acting on each charge is the same. Thus, we here deal
with the inverted with respect to the bottom of the interaction
band distribution of energy levels. Low-energy states of the
electron system correspond to Richardson roots tending to concentrate near
the bottom of the band, while, for the hole system, they
correspond to another distribution of levels, which applies to the
top of the band. It is quite remarkable that these states are
interconnected by various exact relations valid for the Richardson roots.

The electrostatic mapping provides a pictorial representation, which is useful
for our further considerations.

\subsection{Equally-spaced model: arbitrary filling}

Let us now focus on the so-called equally-spaced model, for which
electron and hole electrostatic pictures are characterized by the
same distribution of energy levels, as already have been noted in
Ref. \cite{EPJB}. This model assumes that energy levels $\varepsilon
_{n}^{(e)}$ are located equidistantly between two cutoffs,
$\varepsilon_{F_{0}}$ and $\varepsilon _{F_{0}}+ (L-1)d$, so that
$\varepsilon _{n}^{(e)} = \varepsilon_{F_{0}} + (n-1) d$, where $n$ runs
from 1 to $L$, $d$ being a distance between two
neighboring energy levels.

We analyze a situation, when the total number of electron pairs $M$
in the band is arbitrary. In the usual BCS theory, $M=L/2$ (half filling).

Let us represent $E_{j}^{(e)}$ as
\begin{equation}
E_{j}^{(e)}=2\varepsilon_{F_{0}}+e_{j}^{(M,L)},
\nonumber
\end{equation}
where the notation $(M,L)$ shows that $e_{j}^{(M,L)}$ corresponds to $M$ pairs and $L$ available
states.
These quantities satisfy the set of Richardson equations. The
$j$-th equation ($j=1,\ldots,M$) can be written as:
\begin{equation}
1=\sum_{m=0}^{L-1}\frac{g}{2md-e_{j}^{(M,L)}}+\sum_{l=1 (\neq
j)}^{M}\frac{2g}{e_{j}^{(M,L)}-e_{l}^{(M,L)}}. \label{RichardsonEquallySpaced}
\end{equation}

Eq. (\ref{EHLambdaFinal}) now takes a simple form
($m=0,1,\ldots,L-1$)
\begin{equation}
\sum_{j=1}^{M}\frac{g}{2md-e_{j}^{(M,L)}}+\sum_{j=1}^{L-M}\frac{g}{2[(L-1)-m]d-e_{j}^{(L-M,L)}}=1.
\label{EHLambdaEQS}
\end{equation}
When deriving this equation, we replaced a summation over $m$ in
Richardson equations for holes by a summation over
$[(L-1)-m]$. This trick enabled us to keep a universal
form of Richardson equations and to relate $\{e_{j}^{(M,L)}\}$
($j=1,\ldots,M$) and $\{e_{j}^{(L-M,L)}\}$
($j=1,\ldots,L-M$), which are solutions of the two systems
of equations differing from each other only by their number, the
form of this system (\ref{RichardsonEquallySpaced}) being universal.
This simplification is due to the specifical character of an
energy-level distribution. Of course, it also works for any other distribution,
symmetric with respect to the middle of the interaction band.

Note that in Ref. \cite{EPJB} the Hamiltonian electron-hole symmetry
was used to derive exact relation between the ground state energy of
$M$ and $L-M$ pairs, which are given by the sums of
Richardson roots. Such a relation also follows from Eq. (\ref{eigenehtexplicit}), as we already mentioned.

The Hamiltonian eigenstate in the particle representation can be written as
\begin{equation}
\prod_{j=1}^{M}\mathbb{B}^{(e)\dagger}\left(e_{j}^{(M,L)}\right)|0_{e}\rangle,
\label{eigenstatep}
\end{equation}
where
\begin{equation}
\mathbb{B}^{(e)\dagger}(r)=\sum_{l=0}^{L-1}\frac{B_{l}^{(e)\dagger}}{2ld-r},
\label{Bethevector}
\end{equation}
while $|0_{e}\rangle$ is a vacuum of particle-pairs,
$B_{l}^{(e)}|0_{e}\rangle=0$. The same eigenstate, up to the
numerical factor, can be written in the hole representation as
\begin{equation}
\prod_{j=1}^{L-M}\mathbb{B}^{(h)\dagger}\left(e_{j}^{(L-M,L)}\right)|0_{h}\rangle,
\label{eigenstatephole}
\end{equation}
where
\begin{equation}
\mathbb{B}^{(h)\dagger}(r)=\sum_{l=0}^{L-1}\frac{B_{l}^{(h)\dagger}}{2(L-1-l)d-r},
\label{Bethevectorhole}
\end{equation}
while the vacuum of hole-pairs is $|0_{h}\rangle=B_{0}^{(e)\dagger}
B_{1}^{(e)\dagger} \ldots B_{L-1}^{(e)\dagger} |0_{e}\rangle$.

It is also of interest to explore in more details an impact of
particle-hole symmetry in the thermodynamical limit, for which an
analytical approach based on electrostatic mapping is known, while
the results for the low energy part of the spectrum coincides with
the mean-field results \cite{Gaudin,Rich3,AmicoEl,RomanNucl}. In
this limit, $L\rightarrow \infty$ and $d \rightarrow 0$ in such a
way that $Ld \equiv \Omega $ stays constant, as well as $M/L$ and
$g/d$.  We restrict ourselves to the case of a ground state only. A
fraction of Richardson energies in this case forms complex conjugate
pairs, which are arranged into a single arc, while the remaining
energies are real being located between the lower single-particle
energies. We denote the endpoints of the arc as $\mu \pm \Delta$,
where $ \mu $ and $ \Delta $ are real parameters. Using methods of
electrostatics, it is possible to derive \cite{AmicoEl,RomanNucl}
two  coupled equations for $\mu$ and $\Delta $
\begin{eqnarray}
& \int_{0}^{2\Omega}\frac{d\varepsilon}{\sqrt{(\varepsilon-\mu)^{2}+\Delta^{2}}}=\frac{2d}{g}, \notag \\
& \int_{0}^{\Omega}\frac{\varepsilon-\mu}{\sqrt{(\varepsilon-\mu)^{2}+\Delta^{2}}}d\varepsilon=2d(L-2M).
\nonumber
\end{eqnarray}
These two equations can be solved explicitly yielding
\begin{eqnarray}
& \Delta/4d=\sqrt{M(L-M)}\frac{\exp(-d/g)}{1-\exp(-2d/g)}, \notag \\
& \mu/2d=M-(L-2M)\frac{\exp(-2d/g)}{1-\exp(-2d/g)}.
\label{gapchempotexplic}
\end{eqnarray}
It can be seen that $\Delta$ is nothing but twice the BCS gap,
which characterizes energy difference between the first excited
state and the ground state. For the half filling the above
expression reduces to the usual expression for the gap. At the same
time, $\mu$ is a chemical potential for a particle-pair. For the
half filling, $\mu/2$ falls exactly into the middle of the
interaction band.

We see that there exists a clear manifestation of the particle-hole
symmetry in Eq. (\ref{gapchempotexplic}). Indeed, $\Delta$ stays
invariant under the replacement $M \rightarrow L-M$ which is
expectable due to its role in the excitation spectrum. This
invariance also implies that two arcs for Richardson roots in the
particle and hole representation has endpoints with the same
imaginary parts. We also see that $(\mu^{(e)}+\mu^{(h)})/2d=L$,
which means that $\mu^{(e)}/2$ and $\mu^{(h)}/2$, as well as real
parts for the endpoints of the two arcs are mirror-symmetrical with
respect to the middle of the interaction band. These results for the
energy spectrum are in a full agreement with the treatment of Ref.
\cite{PhysC}, where a mean-field solution of the BCS Hamiltonian for
arbitrary filling $M/L$ was presented, i.e., without utilizing the
Richardson exact solution. Note that for small $M/L$, as well as for
$M/L$ approaching 1, $\Delta$ does not play a role of a minimal
energy difference between the excited states and the ground state.
These limits have similarities with the dilute regimes of Cooper
pairs made of particles and holes, respectively. In these cases, a
role of a minimal excitation energy is played by the binding energy
of a single pair. Thus, changing $M/L$ is analogous to the crossover
between the Bose-Einstein condensate and BCS condensate of fermionic
pairs; for more details see \cite{We,PhysC}.

In the thermodynamical limit, the expression of the condensation
energy in the ground state reads as
\begin{eqnarray}
E_{c}=\frac{\Delta^{2}}{2d}(1-\exp(-2d/g)).
\label{Eground}
\end{eqnarray}
We again see a presence of the particle-hole symmetry in it. Namely,
$E_{c}$ is  proportional both to the particle number and hole
number.

Equation (\ref{EHLambdaFinal}) (or (\ref{RichardsonEquallySpaced}))
implies  that the sum of conserved quantities in the particle and
hole representations, as obtained from the electrostatic mapping in
the thermodynamical limit, must be zero. The values of these
conserved quantities, within such a treatment, have been found in
Ref. \cite{RomanNucl}. The results of Ref. \cite{RomanNucl} are in
agreement with our findings. Indeed, in order to obtain a value of a
conserved quantity at arbitrary filling $M/L$ from it for the half
filling, one has to replace $\varepsilon$ in Eq. (73) of Ref.
\cite{RomanNucl} by $\varepsilon - \mu/2$, where $\mu/2$ is given by
Eq. (\ref{gapchempotexplic}). Within the notations of Ref.
\cite{RomanNucl}, the particle-hole transformation is associated
with the replacements $\varepsilon \rightarrow 2\omega -
\varepsilon$ and $\mu/2 \rightarrow 2\omega - \mu/2$. It is easy to
see that under this transformation, the conserved quantity
$\lambda$, as given by Eq. (73) of Ref. \cite{RomanNucl}, does
change its sign.

\subsection{Equally-spaced model: half-filling}

Perhaps, the most interesting (and most physical) situation is a
half-filling, when $M=L-M$, so that sets $\{e_{j}^{(M,L)}\}$
($j=1,\ldots,M$) and $\{e_{j}^{(L-M,L)}\}$ ($j=1,\ldots,L-M$)
satisfy the same set of  equations. There are multiple solutions for
this set of equations, but an additional constraint exists that the
sums of elements in the sets $\{e_{j}^{(M,L)}\}$ and
$\{e_{j}^{(L-M,L)}\}$ must be the same. Hence, if there is no
degeneracy, these two sets must also coincide.

Thus, the system of equations (\ref{EHLambdaEQS}) is
reduced to
\begin{equation}
\sum_{j=1}^{L/2}\frac{1}{2md-e_{j}^{(L/2,L,\alpha)}}+\sum_{j=1}^{L/2}\frac{1}{2[(L-1)-m]d-e_{j}^{(L/2,L,\beta)}}=1/g,
\label{EHLambdaEQSHalf}
\end{equation}
where $\alpha$ and $\beta$ refer to possibly different solutions of
the same set of equations.  This set of equations is supplemented by
the condition for the energies
\begin{equation}
\sum_{j=1}^{L/2} e_{j}^{(L/2,L,\alpha)}=\sum_{j=1}^{L/2} e_{j}^{(L/2,L,\beta)}.
\label{EHenergies}
\end{equation}

If there is no degeneracy ($\alpha=\beta$), it can be readily seen
that not all of equations (\ref{EHLambdaEQSHalf}) are independent.
Actually, the equations for $m$ and $L-1-m$ are identical.
Therefore, the appropriate values of $m$ are $m=0,1,\ldots,L/2-1$.
Each equation is now related to its own couple of one-electron
energy levels, located mutually symmetrically with respect to the
middle of the interaction band. Besides, each equation contains a
symmetric sum over all rapidities. In this sense, they have a very
different form compared to usual Richardson equations.  Note that in
the most elementary case, $L=2$, Eq. (\ref{EHLambdaEQSHalf}) is
correctly reduced to the single Richardson equation.

We are now going to analyze under which conditions solutions with
$\alpha=\beta$ can exist. It is convenient to classify the solutions
of Richardson equations starting from the limit $g=0$, in which
these solutions must correspond to noninteracting particles. In this
limit each Richardson root approaches one of the single-particle
energy levels. Some of the energy levels thus are occupied, while
the remaining levels are empty. Therefore, each state can be
characterized by the ordered set of occupied (or equivalently empty)
levels. The lowest energy solution corresponds to the first $L/2$
single-particle energy levels occupied. It is convenient to use a pictorial
representation of Ref. \cite{Roman}. Namely, an
occupied and empty levels are depicted as $\bullet$ and $\circ$,
respectively.

It can be readily seen from Eq. (\ref{EHLambdaEQSHalf}) that this
set of  equations has a solution with $\alpha=\beta$ provided that
one level from each couple of levels, located mutually symmetrically
with respect to the middle of the interaction band (Fermi level), is
occupied and another one is empty. For instance, for $L=8, M=4$,
this happens for the state \mbox{$\bullet \bullet \bullet \bullet
\circ \circ \circ \circ$} or \mbox{$\bullet \circ \bullet \bullet
\circ \circ \bullet \circ$} and does not happen for the state
\mbox{$\circ \bullet \bullet \bullet \bullet \circ \circ \circ$}.
Indeed, if the above condition is not satisfied than there should be
at least one couple of such levels, for which both these levels are
empty. The equation (\ref{EHLambdaEQSHalf}) for this couple has no
solution in the limit $g \rightarrow 0$, since the right-hand side
of this equation contains a singular term $1/g$, while the left-hand
side is not divergent. Hence, not all Richardson solutions belong to
this class of states. It is easy to realize that the number of such
states is $2^{L/2}$, while the total number of solutions is known to
be $\binom{L}{L/2}$. We will call the states, for which the solution
of (\ref{EHLambdaEQSHalf}) with $\alpha=\beta$ exists, self-dual
states, since for them Hamiltonian eigenstates both in the particle
and hole representation are expressed through the same set of
rapidities, see Eqs. (\ref{eigenstatep}, (\ref{Bethevector}),
(\ref{eigenstatephole}, (\ref{Bethevectorhole}.

If some state is not self-dual, than there should be another
solution for which the whole set of equations
(\ref{EHLambdaEQSHalf}) is satisfied together with the constraint
(\ref{EHenergies}), i.e., there must be solutions with  $\alpha \neq
\beta$. It is again easy to see that such solutions do exist. They
correspond to states, which are obtained from each other by a mirror
reflection with respect to the Fermi level accompanied by the mutual
replacement $\bullet \leftrightarrow \circ$. This is again evident
from equations (\ref{EHLambdaEQSHalf}), in which singularities in
their right-hand sides are compensated by singularities in the
left-hand sides, while (\ref{EHenergies}) is fulfilled
automatically. For example, the states \mbox{$\circ \bullet \bullet
\bullet \bullet \circ \circ \circ$} and \mbox{$ \bullet \bullet
\bullet \circ \circ \circ \circ \bullet $} are dual to each other.
Thus, any state in $g \rightarrow 0$ limit is either self-dual or
there is another solution, which is dual to this state and has the
same energy. Actually, the ground state is always self-dual, as well
as a first excited state. Note that in the limit $L \rightarrow
\infty$, the total number of solutions scales as $\binom{L}{L/2}
\sim 2^{L}$, so that the fraction of self-dual states is
exponentially small.

\begin{figure}[h]
\center\includegraphics[width=0.35\linewidth]{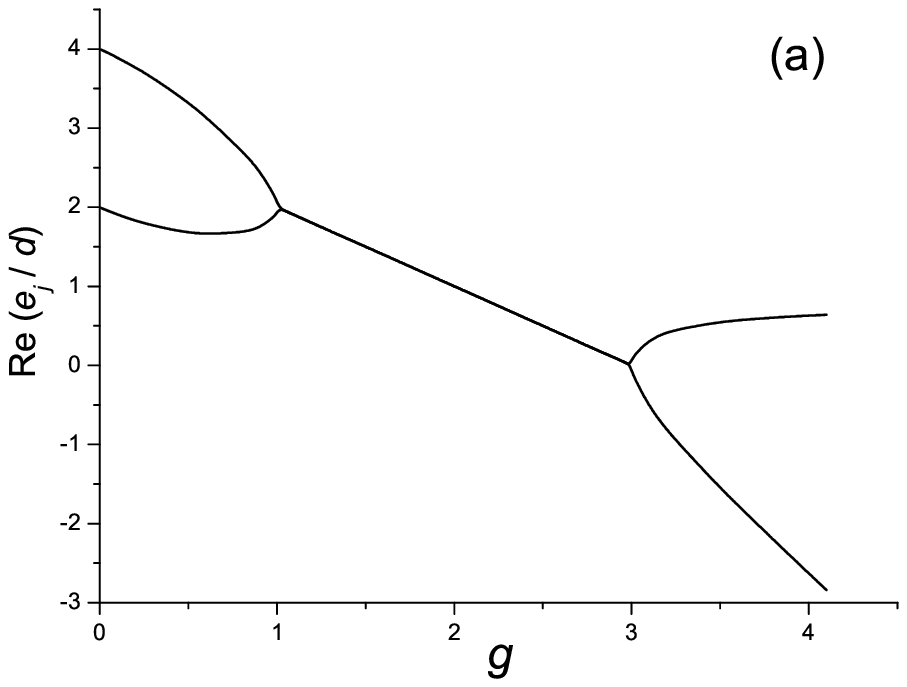}
\center\includegraphics[width=0.35\linewidth]{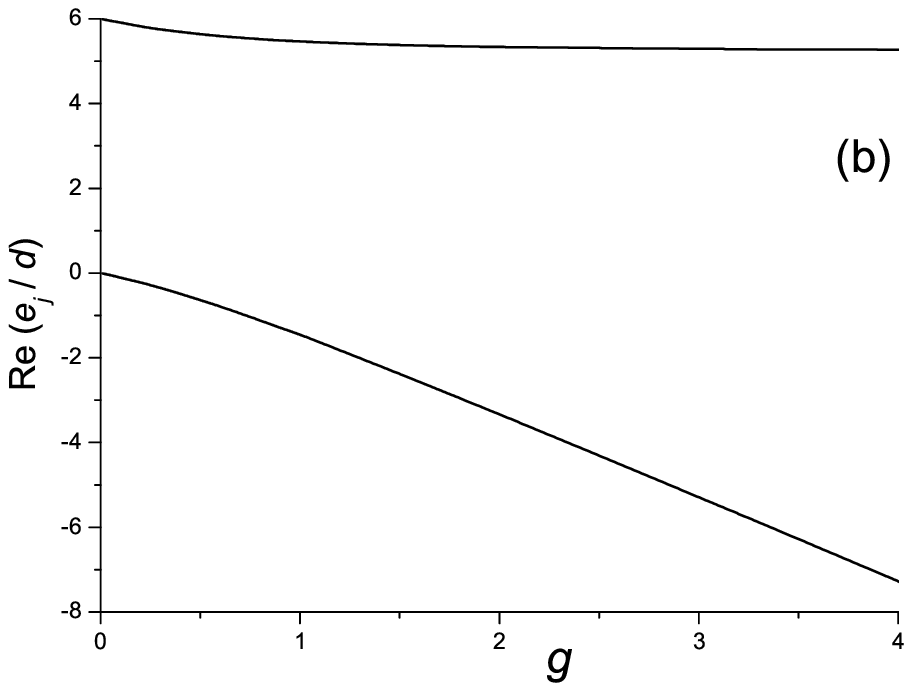} \caption{
Real parts of two sets of Richardson roots corresponding to mutually
dual states as a function of $g$ at $L=4, M=2$. These two states are
\mbox{$\circ \bullet \bullet \circ$} (a) and \mbox{$ \bullet \circ
\circ \bullet $} (b)}.
\end{figure}

It is not evident at all what is going to happen with this
classification outside of a quite specific limit $g \rightarrow 0$,
when Richardson roots decouple from one-particle energy levels and
start very peculiar transformations, which can also include highly
singular points. To clarify this issue, we solved Richardson
equations numerically. We restricted ourselves to configurations
with few pairs only, because the number of solutions grows
exponentially with $L$, while we are interested in the total set of
all possible solutions to see connections between them.

Our general conclusion is that the classification appearing in $g
\rightarrow 0$  limit stays universal. Namely, self-dual states does
not change their character, so that their Richardson roots satisfy
(\ref{EHLambdaEQSHalf}) with $\alpha=\beta$. Dual states also stay
internally coupled through (\ref{EHLambdaEQSHalf}) and
(\ref{EHenergies}) with $\alpha \neq \beta$ along the whole
crossover from $g \rightarrow 0$ to $g \rightarrow \infty$ limit. We
would like to stress that this coupling is far from being trivial,
since Richardson roots for two given dual states are not the same.
Moreover, they can follow very different transformations, so that
even numbers of singularities passed as $g$ increases, can be
different. This is illustrated by Figures 1 and 2, which provide
dependencies of Re $E_j$ as functions of $g$ for two mutually dual
states corresponding to $L=4, M=2$ (Fig. 1) and $L=6, M=3$ (Fig. 2).
The dual states in the first case are \mbox{$\circ \bullet \bullet
\circ$} and \mbox{$ \bullet \circ \circ \bullet $}, while in the
second case they are \mbox{$ \bullet \bullet \circ \circ \circ
\bullet$} and \mbox{$  \circ \bullet \bullet \bullet \circ \circ $}.
It is seen from Fig. 1a that there are two singularities for the
state \mbox{$\circ \bullet \bullet \circ$}, while there is no
singularity for a dual state \mbox{$ \bullet \circ \circ \bullet $}.
For each of the states depicted in Fig. 2a and 2b, there is only one
singularity, but in appears at slightly different values of $g$.

\begin{figure}[h]
\center\includegraphics[width=0.35\linewidth]{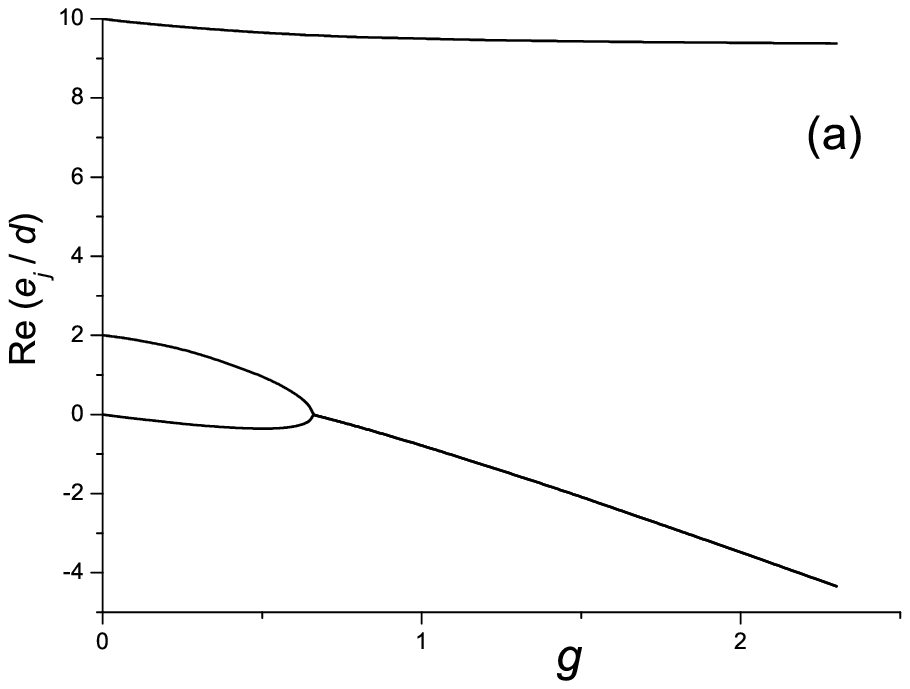}
\center\includegraphics[width=0.35\linewidth]{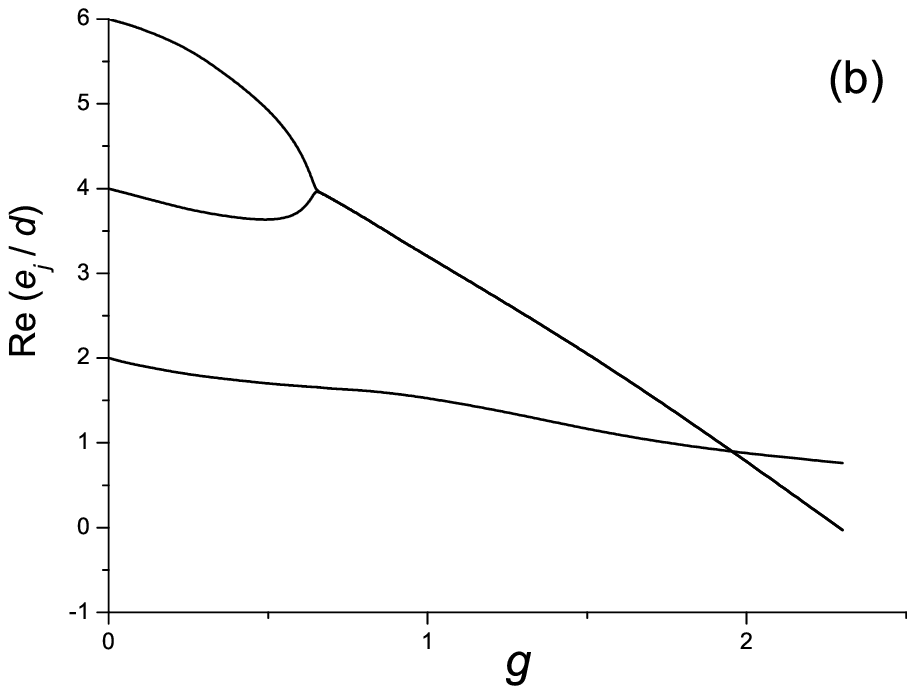} \caption{
Real parts of two sets of Richardson roots corresponding to mutually
dual states as a function of $g$ at $L=6, M=3$. These two states are
\mbox{$ \bullet \bullet \circ \circ \circ \bullet$} (a) and \mbox{$
\circ \bullet \bullet \bullet \circ \circ $} (b)} .
\end{figure}

We now provide the full 'map' of the states for $L=4, M=2$ and $L=6,
M=3$. The self-dual solutions for $L=4, M=2$ are: \mbox{$ \bullet
\bullet \circ \circ $}, \mbox{$ \bullet \circ \bullet \circ $},
\mbox{$ \circ \bullet \circ \bullet$}, and \mbox{$ \circ \circ
\bullet \bullet$}, while mutually dual solutions are \mbox{$\circ
\bullet \bullet \circ$} and \mbox{$ \bullet \circ \circ \bullet $}.
The self-dual solutions for $L=6, M=3$ are: \mbox{$ \bullet \bullet
\bullet \circ \circ \circ$}, \mbox{$ \bullet \bullet \circ \bullet
\circ \circ$}, \mbox{$ \bullet \circ \bullet \circ  \bullet \circ$},
\mbox{$ \bullet \circ \circ \bullet \bullet \circ$}, \mbox{$ \circ
\bullet \bullet \circ \circ \bullet$}, \mbox{$ \circ \bullet \circ
\bullet \circ \bullet$}, \mbox{$ \circ \circ \bullet \circ \bullet
\bullet$}, \mbox{$ \circ \circ \circ \bullet \bullet \bullet$}. The
couples of mutually dual states are \mbox{$ \bullet \bullet \circ
\circ \bullet \circ$} and \mbox{$ \bullet \circ \bullet \bullet
\circ \circ$},
 \mbox{$ \bullet \bullet \circ \circ \circ \bullet $} and \mbox{$ \circ \bullet \bullet \bullet \circ \circ$},
\mbox{$ \bullet \circ \bullet \circ \circ \bullet $} and \mbox{$
\circ \bullet \bullet \circ \bullet \circ $}, \mbox{$ \bullet \circ
\circ \bullet \circ \bullet $} and \mbox{$ \circ \bullet \circ
\bullet \bullet \circ $}, \mbox{$ \bullet \circ \circ \circ \bullet
\bullet $} and \mbox{$ \circ \circ \bullet \bullet \bullet \circ $},
\mbox{$ \circ \circ \bullet \bullet \circ \bullet $} and \mbox{$
\circ \bullet \circ \circ  \bullet \bullet $}.  Note that in the
case $L=2, M=1$ both possible states, \mbox{$ \bullet \circ $} and
\mbox{$ \circ \bullet $}, are self-dual. It is of interest that each
self-dual solution has a 'partner', which is also self-dual and can
be obtained by the replacement $ \bullet \leftrightarrow  \circ $
(not accompanied by the mirror reflection with respect to the Fermi
level, as in a duality of the first type).

Another important conclusion is that the suggested classification
also works for any other distribution of levels,  mirror-symmetric
with respect to the Fermi level. This again can be readily seen in
$g=0$ limit. To verify it outside of this limit, we used numerics
for various distributions of energy levels. Note that, in this case,
it is of course necessary to modify (\ref{EHLambdaEQSHalf}) by
inserting actual distribution of one-particle energies instead of
the equally-spaced set.

Thus, relations (\ref{EHLambdaEQSHalf}) form a system of equations,
which are complementary and equivalent to the system
of Richardson equations at half-filling and for the equally-spaced
model (and after a minimal modification for  any other
mirror-symmetric with respect to the Fermi level distributions of
one-particle energies). Together with Richardson equations, the
relations (\ref{EHLambdaEQSHalf}) form an overdetermined set of
equations. Hopefully, these equations may provide additional tools
for the resolution of the problem or computing correlation
functions. Notice that relations (\ref{EHLambdaEQSHalf}) lack
singular terms of the form $1/(E_{1}-E_{2})$ typical for Richardson
equations. This feature seems to be attractive for purposes of
numerical calculations, since the above classification provides
simple tools to understand if a given state in $g \rightarrow 0$
limit is self-dual or not. If it is not self-dual than it is
straightforward to determine a solution, which is dual to it in the
same limit. Once the initial configuration is known, one can solve
the set of equations (\ref{EHLambdaEQSHalf}) for Richardson roots
for the two states, as $g$ increases.

Our approach can be related to a recent development
\cite{Gritsev,Ghent} on solving the  Richardson equations in terms
of quantities $\Lambda_m$ instead of rapidities, which can be
defined as
\begin{equation}
\Lambda_m=\sum_{j=1}^{M}\frac{1}{2\varepsilon
_{m}^{(e)}-E_{j}^{(e)}}. \label{Lambda}
\end{equation}
These quantities satisfy the system of equations which turns out to be independent on $M$ for a given $L$  \cite{Gritsev,Ghent}. An additional constraint\cite{Gritsev,Ghent} is used to extract solutions relevant to a given $M$. Actually, we here show that for self-dual solutions this system of equations can be significantly simplified, since for them $\Lambda_m+ \Lambda_{L-1-m} = 1/g$. For the couples of mutually-dual solutions $\Lambda_m$ and $\Lambda_{L-1-m}$, connected by this identity, correspond to different solutions of the same set of equations.

Note that our results might be also of interest for pure mathematics, since
they deal with somehow hidden relations between solutions of two
sets of polynomial equations.

\section{Conclusions}

A generalized Richardson model also known as Russian doll BCS model, on the level of Hamiltonian, is characterized by the electron-hole pairing duality. We examined consequences of this duality on the solution of the model by the algebraic Bethe ansatz. We pointed out that, within the both representations, the model is integrable, but the number of rapidities in these two pictures are, in general case, different. In addition, bare kinetic energies of particle-pairs and hole-pairs are inverted with respect to each other. Nevertheless, both representations should provide the same eigenvalues for quantum invariants.

By analyzing these quantum invariants, we obtained a set of quite nontrivial relations between the Bethe roots in the electron and hole representations. Together with the two initial sets of Bethe equations, they form an overdetermined system of equations. In the quasi-classical limit, similar relations found in Ref. \cite{Faribault} for Gaudin models are recovered from our result.
We point out that these relations take the most simple form at half-filling and at some special realizations of one-body energy-level distributions (symmetric with respect to the middle of the interaction band including a well known equally spaced model). In this case, the sets of Richardson roots in the electron and hole representations must be the same up to the translation. We analyzed how different solutions of Richardson equations are linked under the transformation from the particle representation to the hole representation. We have shown that there are two types of solutions. The solutions of the first type turn into themselves (in particular, the ground state solution), while the solutions of the second type appear by couples, so that they transform to each other.

\section{Acknowledgments}

Useful discussions with A. S. Gorsky are acknowledged. W. V. P. also
acknowledges discussions with M. Combescot, as well as a support
from the RFBR (project nos. 12-02-00339 and 15-02-02128), RFBR-CNRS programme
(project no. 12-02-91055), and the Russian Science Support
Foundation. L. V. B. is grateful for the support from the RFBR
(project no. 14-02-00494).

\end{document}